\documentclass[a4paper,12pt,twocolumn]{article}
\usepackage{cite}
\usepackage[dvips,hiresbb]{graphicx}
\renewcommand{\figurename}{Fig.}

\title{\textbf{Special polarization characteristic features of a three-dimensional terahertz photonic crystal semi-quantitatively evaluated by using a FEM}}
\author{\Large{Chikara Sakurai} \thanks{Email: c-sakurai@river-ele.co.jp (primary), sakuraikazan@gmail.com (second)} \\ \textit{River Electric Corporation, 2-1-11 Fujimigaoka, Nirasaki, Yamanashi, Japan}} 
\date{\today}
\begin{document}
\onecolumn{
\maketitle
\begin{abstract}
\normalsize{A previous work~\cite{sakurai1} experimentally confirmed that the special polarization characteristic features of a three-dimensional terahertz (THz) photonic crystal with a silicon inverse diamond structure whose lattice point shape was vacant regular octahedrons, did not apply to general physical and optical basic rules in appearance. 
One of the basic rules is that, according to Maxwell's equations, the electric-field  direction of the reflected wave rotates by 180 degrees from that of the incident wave in the case of complete reflection (Bragg reflection) and normal incidence.
It is said that three-dimensional photonic crystals have no polarization anisotropy within photonic band gap (stop gap, stop band) of high symmetry points in normal incidence.  Experimental results, however, confirmed that the polarization orientation of a reflected wave was rotated by 90 degrees for that of an incident wave at a frequency within photonic band gap.
In this work, measured reflected spectra~~\cite{sakurai1} were semi-quantitatively evaluated by being compared with reflection ones by using a FEM (finite element method). These results suggest that measured polarization anisotropy does not apply to Maxwell's equations in appearance but applies to these equations in essential as expected.} 
\end{abstract}

\twocolumn
\section*{Introduction}
\hspace*{5mm}The previous work, ref.~\cite{sakurai1} \footnote{ Ref.~\cite{sakurai1} contains more detailed analyses than those in ref.~\cite{sakurai2}.} with~the polarization anisotropy, which was the polarization orientation (electric-field direction) difference between the reflected wave and the incident one, confirmed~that~new reflection phenomena of a three-dimensional (3D) terahertz (THz) photonic crystal (PC) with a silicon inverse diamond structure whose lattice point shape was vacant regular octahedrons, did not apply to general physical- and optical- basic rules in appearance.\\
\hspace*{5mm}These basic rules are as follows. \\
\hspace*{5mm}(A) Electric-field is vector, and various optical phenomena are analyzed by using methods of resolution and synthesis of electric-field.\\
\hspace*{5mm}(B) According to Maxwell's equations that are electromagnetic basic rules, 
electric-field direction of the reflected wave rotates by 180 degrees from that of the incident wave in the case of complete reflection (Bragg reflection) and normal incidence.\\
\hspace*{5mm}(C) According to crystallography, diamond structure is in cubic system and optically isotropic.\\
\hspace*{5mm}Polarization features in ref.~\cite{min} do not also apply to rule (B) except complete reflection and rule (C).
The features exist in frequencies without photonic band gap (PBG).
The 3D-PC has eigen modes (branches) in these frequencies. Each eigen mode has an intrinsic symmetry, and some polarization anisotropy also exists.\\
\hspace*{5mm}In ref.~\cite{pri}, the direction- and wavelength-dependent polarization anisotropy of stop gap branching has been studied with fcc symmetry as a function of an incidence angle for two polarization orientation normal to each other, TE and TM.\\
\hspace*{5mm}In the previous work~\cite{sakurai1}, however, the polarization anisotropy in normal incidence and within X point's PBG, BGX in fig.~A.\ref{fig:bandstructure} (appendix), was entirely different from above characteristics; no eigen modes based on the 3D-band structure exist within BGX. None the less because some two directions at right angles to each other on the plane (001) are constitutionally and optically identical, the polarization orientation of the reflected wave was rotated by 90 degrees from that of the incident wave at a frequency within BGX when that of the incident wave was the direction that bisects these two directions.\\
\hspace*{5mm}The general outline of ref.~\cite{sakurai1} is shown in fig.~\ref{fig:polar}. The comparison of  the polarization characteristic features as generally expected 
in fig.~\ref{fig:polar}(a) with ones in the measured spectra in fig.~\ref{fig:polar}(c) is shown. I(X,Y) in fig.~\ref{fig:polar}(b) is the polarization orientation of the incident wave. The unit cell of a diamond structure in fig.~\ref{fig:latticeA} is illustrated. The incident wave is normal incidence, [001] ($\Gamma$-X direction) \footnote{[001] and (001) are defined on the X-Y-Z coordinate system in fig.~\ref{fig:latticeA}.}. The solid arrow is the polarization orientation of the incident wave, and the dashed arrow is that of the reflected wave. The frequency of the incident wave is 0.42 THz within BGX.\\
\hspace*{5mm}General characteristic features of the polarization indicate that the polarization orientation of the reflected wave is parallel to that of the incident wave as shown in fig.~\ref{fig:polar}(a). The  polarization features in the measured spectra, however,  indicated entirely different ones as shown in fig.~\ref{fig:polar}(c); the polarization orientation of the reflected wave was perpendicular to that of the incident wave when that of  the incident wave was [I($\pm 1$, 0)] or  [I(0, $\pm 1$)].\\
\begin{figure}[t]
\centering
\includegraphics[width=7.8cm,]{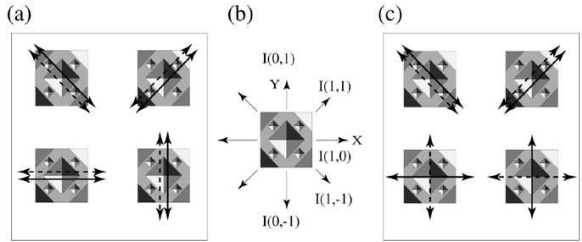}
\caption{\footnotesize Incident wave is normal incidence, [001]. The heavy solid arrow is the polarization orientation of the incident wave, and the heavy dashed arrow is that of the reflected wave. (a) Features expected from general basic optical rules. (b) I(X,Y) is the polarization orientation of the incident wave.  (c) Special characteristic features of measured spectra~\cite{sakurai1}.}
\label{fig:polar}
\end{figure}
\hspace*{5mm}In contrast, when that of the incident wave is [I($\pm 1$, $\pm 1$)] or [I($\pm 1$, $\mp 1$)] (double sign in same order), no polarization anisotropy existed experimentally \footnote{In the experiment, reflectivity and transmittance were only measured.
In the later FEM (finite element method) analyses (arXiv:1811.02990v2), it was found that phase-difference existed in spite of diamond structure. The reason has still remained unsolved.}.
In other words, the polarization orientation of the reflected wave was parallel to that of the incident wave. According to rule (A), when that of the incident wave is [I($\pm 1$, 0)] or [I(0$, \pm 1$)], the polarization orientation of the reflected wave is parallel to that of the incident wave. The experimental facts, however, indicated that the former was perpendicular to the latter.
These facts do not apply to rule (A) in appearance, in addition, the characteristic features do not apply to the general theory of the 1/2 wave plate \footnote{In this theory, the two directions orthogonal each other are optically anisotropic.} in appearance.\\
\hspace*{5mm}In this work, reflected and transmission spectra of the 3D-PC were analyzed by using a FEM (finite element method), and these simulation spectra were semi-quantitatively made comparisons with the measured spectra.
\section*{Parameters}
\hspace*{5mm}The FEM software was Femtet(R)2017.1 made in Murata Software Co., Ltd..   The model size in this work was from 1500 $\mu$m to 1800 $\mu$m on a side.  Several models were analyzed with the size magnified a hundred times since the maximum drawing size was 1000 $\mu$m and 1000 mm when analysis unit was $\mu$m and mm, respectively. The scaling rule is applicable in the photonic crystals. In other words,
frequency $\times$ size (lattice constant) is constant. On the base of this rule, the spectrum frequencies were finally shifted from GHz to 0.1 $\times$ THz as the size needed to be reduced by a hundredth. The following explanations use final sizes ($\mu$m) and final frequencies (THz).\\ 
\hspace*{5mm}In addition, the software had no command to obtain S-polarization (S-p) spectra and P-polarization (P-p) ones separately. Only compound spectra of S-p and P-p \footnote{S-p and P-p are orthogonal each other. Two polarization are explained in detail in fig.~A.\ref{fig:RTsystem}.} were able to be obtained.\\
\begin{figure}[t]
\centering
\includegraphics[width=6.5cm,]{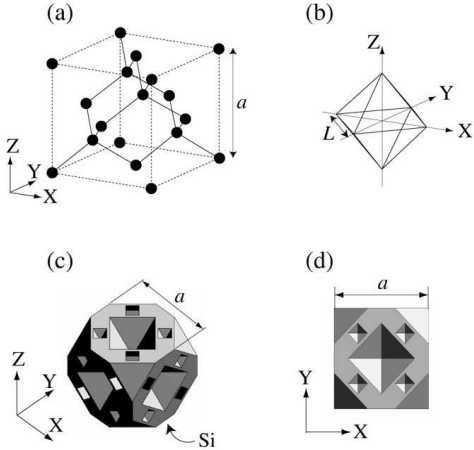}
\caption{\footnotesize (a) Lattice of the diamond structure. (b) Shape of the lattice point is regular octahedrons. It is vacant. The surrounding material is Si. (c), (d) Unit cell of the Si inverse diamond structure.}
\label{fig:latticeA}
\end{figure}
\hspace*{5mm}The unit cell of the simulation model is the diamond structure as shown 
in fig.~\ref{fig:latticeA}(a). The sphere is the lattice point and its shape is the regular octahedrons in fig.~\ref{fig:latticeA}(b). 
It is vacant (atmosphere) and the dielectric constant, $\varepsilon_1$ = 1.00 was set. 
The surrounding material is pure Si and the dielectric constant, $\varepsilon_2$ = 11.9 was set.
The lattice constant, $a$ = 300 $\mu$m and the side length of the regular octahedrons, $L$ = 150 $\mu$m were set in the theoretical and experimental works.  In the unit cell of the Si inverse diamond structure, many sections of the air octahedrons on the lattice points are arranged as shown in figs.~\ref{fig:latticeA}(c) and \ref{fig:latticeA}(d). \\   
\hspace*{5mm}Fig.~A.\ref{fig:bandstructure}(a) shows the calculated photonic band structure by using plane wave expansion method.
In this work, the direction of the incident wave was  +Z-direction, [001]
in the real space and it corresponds to $\Gamma$-X direction on the $\mathrm{K}_\mathrm{z}$-axis in the wave number space (K-space). 
The polarization anisotropy was studied on the surface (001) at around BGX. \\
\hspace*{5mm}The relationship of reflectivity $R$(XY), transmittance $T$(XY) with the polarization orientations of the incident wave, I(XY) is explained as follows. Four orientations of I(0, $\pm 1$) and I($\pm 1$, 0) are constitutionally and optically comparable as shown in fig.~\ref{fig:polar}(b). they are totally expressed as I(10). For the same reason, four orientations, I($\pm 1$, $\pm 1$) are also totally expressed as I(11) \footnote{According to the later arXiv:1811.02990v2, [I($1, -1$) and I($-1, 1$)] need to be
distinguished from [I(1, 1) and I($-1, -1$)]. In this work, I(11) corresponds to [I($1, -1$) and I($-1, 1$)].}.
The reflectivity, $R$ is expressed as $R$(10) or $R$(11) when the incident wave is I(10) or I(11), respectively, and the transmittance, $T$ is similarly expressed as $T$(10) or $T$(11) when the incident wave is I(10) or I(11), respectively.\\
\hspace*{5mm}Furthermore, both $R$(10) and $R$(11) are compound spectra of S-p and P-p of the reflected wave.
$T$(10) and $T$(11) are also the same in meaning \footnote{$R(f)=RS(f)+RP(f)$. $R(f)$ is $R$(10) or $R$(11). $T(f)=TS(f)+TP(f)$. $T(f)$ is $T$(11) or $T$(10). The variable, $f$ is a THz frequency. $RS$ ($TS$) is S-p reflectivity (transmittance). $RP$ ($TP$) is P-p reflectivity (transmittance). Additionally, for example, $R$(10) is $R(10)(f)$, exactly.}.\\
\hspace*{5mm}In more detail, firstly, (10) and (11) correspond to the polarization orientation of the incident wave, secondly, S-p of the reflected wave means the polarization component parallel to the polarization orientation of the incident wave, thirdly, P-p of the reflected wave means the polarization component perpendicular to S-p.\\
\begin{figure}[t]
\centering
\includegraphics[width=7.5cm,]{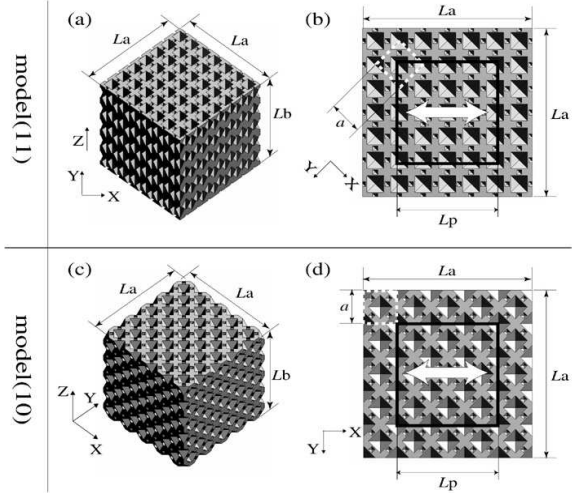}
\caption{\footnotesize Two analyzed models are model(11) and model(10). The square surrounded with black heavy lines, port is the area of the incident wave. The white heavy arrows are the orientation of the electric-field vector of the incident wave.}
\label{fig:latticeB}
\end{figure}
\hspace*{5mm}Two analyzed models are model(11) and model(10) as shown in 
fig.~\ref{fig:latticeB}. X-Y-Z coordinate system corresponds to that in fig.~\ref{fig:latticeA}.
The model size is $L_\mathrm{a}$ $\times$ $L_\mathrm{a}$ $\times$ $L_\mathrm{b}$ ($\mu$m)$^3$. $L_\mathrm{b}$ is the height of the Z-direction. $L_\mathrm{a}$ $\times$
$L_\mathrm{a}$ is the area in X-Y plane.
The square area surrounded  with white dashed lines is $a$ $\times$ $a$ ($\mu$m)$^2$, which corresponds to that in fig.~\ref{fig:latticeA}(d). 
“port” is shown as the square surrounded  with black heavy lines.  It is the area of the incident wave, whose size is $L_\mathrm{p}$ $\times$ $L_\mathrm{p}$ ($\mu$m)$^2$.
The white heavy arrow is the orientation of the electric-field vector, $E_\mathrm{I}$ of the incident wave, which is fixed. The direction of the incident wave is +Z-direction and normal incidence. The location correlation between I(XY) and model(XY) is as follows. \\
\begin{center}
\begin{tabular}{clr}
model(11) : & $E_\mathrm{I}$ $\|$ $(Y=-X)$& ; I(11)\\
model(10) : & $E_\mathrm{I}$ $\|$ $X$ & ; I(10)
\end{tabular}
\end{center}
\begin{table}[t]
\centering
\caption{\footnotesize Five kinds of 3D-PC size and port size. $R$i is reflectivity and $T$i is transmittance}
\includegraphics[width=6cm, trim=0  0 0 -20]{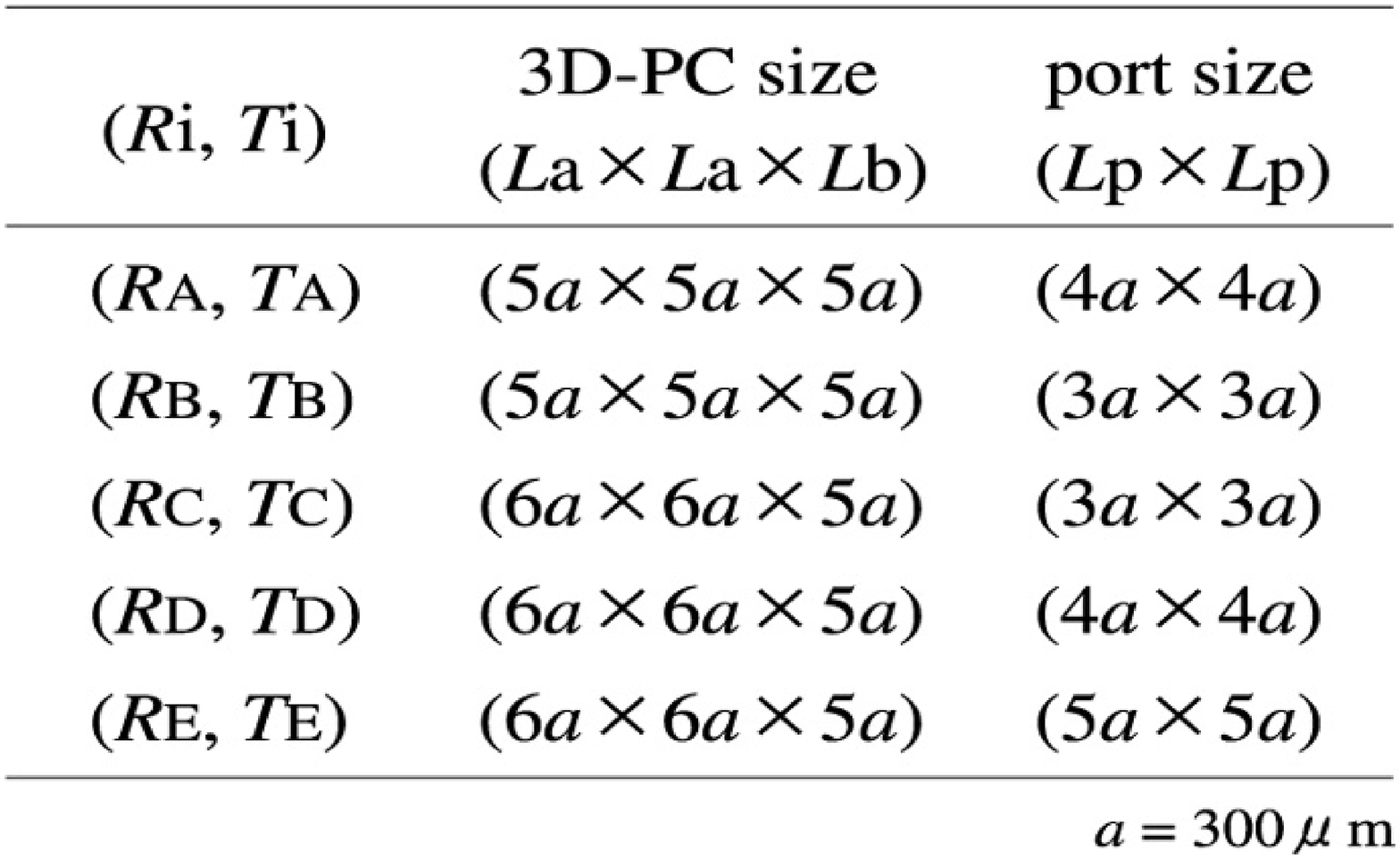}
\label{table:table}
\end{table}
\hspace*{5mm}Table.~\ref{table:table} \footnote{The model size is nearly maximum for~the limit of the personal computer's performance: OS: Windows~10~Home (64bit), processor: Intel(R) Core (TM) i5-6500 CPU@3.20 GHz, RAM: 64.0 GHz.} shows five kinds of (3D-PC size, port size) and (reflectivity, transmittance) abbreviation simulated. Models' mesh size was 
50 $\mu$m. 
\section*{Simulation Results}
\hspace*{5mm}Simulation results of reflectivity ($R_\mathrm{i}$, closed circle) and transmittance ($T_\mathrm{i}$, open circle)
\footnote{Four S-parameters were calculated in the FEM software. They are $S_{11}$,  $S_{22}$, $S_{21}$ and $S_{12}$. Reflectivity and transmittance correspond to $|S_{11}|^2$
and  $|S_{21}|^2$, respectively.} are shown in fig.~\ref{fig:RTABC}. A, B, and C are slected as “i”. The frequency interval of the data was 0.005 THz.~The~absorption~coefficient (cm$^-$$^1$) was set as zero. The value in parentheses on the extreme left is (3D-PC size, port size) in table~\ref{table:table}.\\   
\hspace*{5mm}PBG's in fig.~\ref{fig:RTABC} are nearly equal to BGX in fig.~A.\ref{fig:bandstructure}.
The reflectivity is large and transmittance is very small within PBG.
In model(11) [figs.~\ref{fig:RTABC}(a), \ref{fig:RTABC}(c) and \ref{fig:RTABC}(e)], the shapes of three reflected spectra are approximately convex downward within PBG, which is similar characteristic to one another.
In model(10) [figs.~\ref{fig:RTABC}(b), \ref{fig:RTABC}(d) and \ref{fig:RTABC}(f)], the shapes of the reflected spectra display concavity and convexity within PBG.
The shapes of the spectra depend on the 3D-PC size and port size since these size are small.\\
\begin{figure}[t]
\centering
\includegraphics[width=7.0cm,]{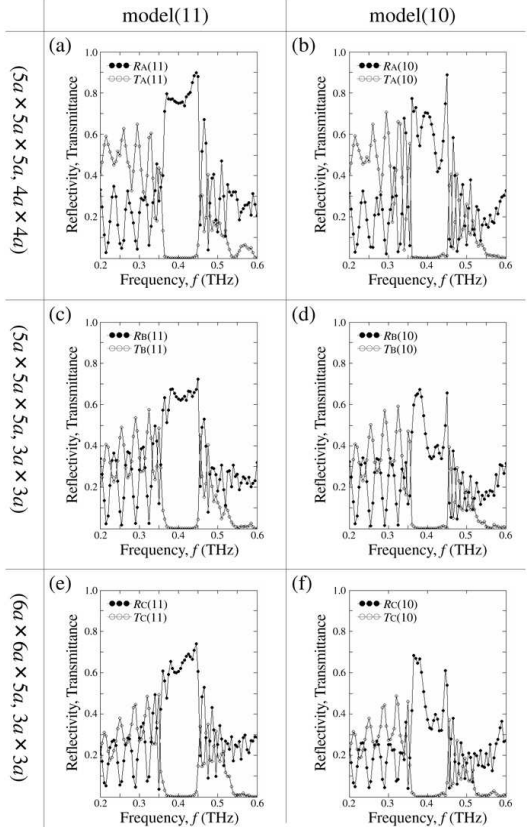}
\caption{\footnotesize Simulation results of two models $\times$ three models' sizes.
The value in parentheses on the extreme left is (3D-PC size, port size).
 In model(11), the shapes of three reflected spectra are approximately convex downward within PBG. In model(10), all three reflected spectra display concavity and convexity within PBG.}        
\label{fig:RTABC}
\end{figure}
\hspace*{5mm}In the case of the same port size, ($R_\mathrm{i}$, $T_\mathrm{i}$) (i = B and C), the spectra are similar each other in model(10).
$R_\mathrm{D}$ was also similar spectra to $R_\mathrm{A}$.
As the port size is larger, the reflectivity within PBG is larger.
When the port size is  5$a$ $\times$ 5$a$, two local maximum values of $R_\mathrm{E}$(10) are 0.82 and 0.93 as shown in fig.~\ref{fig:RTE}.  It is thought that as the port size becomes large, the ratio of a scattering wave to the reflected wave becomes small and then the reflectivity becomes large. The shapes of the reflected spectra in PBG also change to a certain degree.\\
\hspace*{5mm}In the sub-peaks of reflected spectra and transmission ones without PBG in figs.~\ref{fig:RTABC}(a) to \ref{fig:RTABC}(f), local maximum value and local minimum one of two spectra, $R_\mathrm{i}$ and $T_\mathrm{i}$ tend to appear at the same frequency as expected.\\
\begin{figure}[t]
\centering
\includegraphics[width=3.0cm, trim=105 0 0 0]{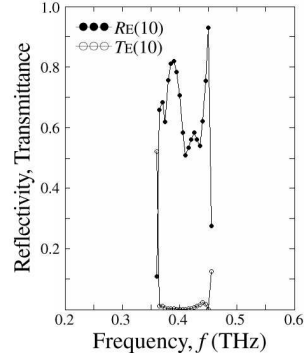}
\caption{\footnotesize (3D-PC size, port size) is (6$a$ $\times$ 6$a$ $\times$ 5$a$, 5$a$ $\times$ 5$a$). $R_\mathrm{E}$(10) and $T_\mathrm{E}$(10) are plotted from 0.355 to 0.450 THz. These values are nearly equal to BGX. As the port size is larger, reflectivity within PBG is also larger.}
\label{fig:RTE}
\end{figure}
\section*{Experimental results}
\hspace*{5mm}The measured reflected spectra, $RS$ and $RP$
and the total reflected ones obtained from them, in ref.~\cite{sakurai1}, are shown in fig.~\ref{fig:R10R11}.   
The total number of layered chips of the sample was 28. The height of one chip is 75 $\mu$m, and the height of the sample corresponds to 7$a$. The reflection and transmission measurement systems are shown in fig.~A.\ref{fig:RTsystem}.\\
\hspace*{5mm}Fig.~\ref{fig:R10R11}(a) shows $RP$(11) (open circle), $RS$(11) (cross shape) and $R$(11)
 (closed circle). $R(11)=RP(11)+RS(11)$. The polarization orientation of the incident wave is I(11), which corresponds to the white heavy arrow in fig.~\ref{fig:latticeB}(b). 
90$^\circ$ (degrees) -spectrum (open circle) in fig.~A.\ref{fig:RSRPRW}(a) corresponds to $RS$(11), and 90$^\circ$-spectrum in fig.~A.\ref{fig:RSRPRW}(b) corresponds to $RP$(11). 
$RP$(11) is very small, and $R$(11) is nearly equal to $RS$(11).\\
\hspace*{5mm}Fig.~\ref{fig:R10R11}(b) shows $RP$(10) (open circle), $RS$(10) (cross shape) and $R$(10)
 (closed circle).  $R(10)=RP(10)+RS(10)$. The polarization orientation of the incident wave is I(10) , which corresponds to the white heavy arrow in fig.~\ref{fig:latticeB}(d). 45$^\circ$-spectra (closed circle) in figs.~A\ref{fig:RSRPRW}(a) and A\ref{fig:RSRPRW}(b) correspond to $RS$(10) and $RP$(10), respectively. At 0.42THz, $RS$(10) is very small, and $R$(10) is nearly equal to $RP$(10).\\
\begin{figure}[t]
\centering
\includegraphics[width=7.8cm,]{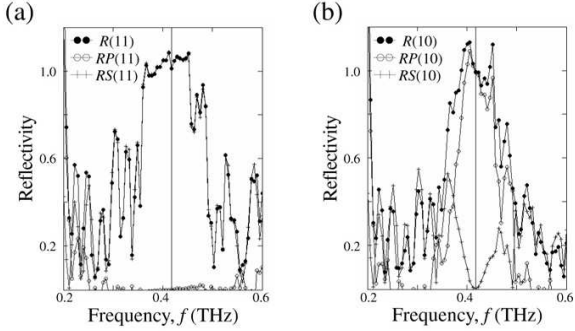}
\caption{\footnotesize $R$(11) and $R$(10) are compound reflected spectra. $RS$(11) and $RS$(10) are measured S-p reflected spectra. $RP$(11) and $RP$(10) are measured P-p ones. $R(11)=RP(11)+RS(11)$.  $R(10)=RP(10)+RS(10)$.}
\label{fig:R10R11}
\end{figure}
\begin{figure}[t]
\centering
\includegraphics[width=7.8cm,]{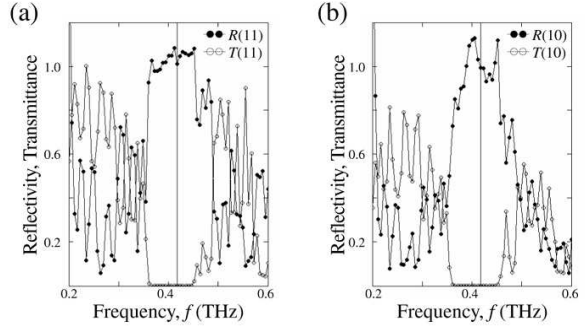}
\caption{\footnotesize (a) Measured compound (S-p + P-p) reflected spectra (closed circle), $R$(11) and transmission ones (open circle), $T$(11). (b) Reflected spectra, $R$(10) and transmission ones, $T$(10), similarly,  }
\label{fig:RT10RT11}
\end{figure}
\hspace*{5mm}Fig.~\ref{fig:RT10RT11} shows $R$(11), $T$(11),  $R$(10) and $T$(10).
$T(11)= TS(11)+TP(11)$.  $T(10)= TS(10)+TP(10)$.
$T$(11) and $T$(10) are nearly equal to zero within BGX.\\
\hspace*{5mm}The number of side peaks without BGX depends on the layered number that corresponds to the height of Z-direction. The height of the total layered chips, 7$a$ is larger than 5$a$ that is the height of the simulation model. The number of side peaks in the measured spectra is larger than those in the simulation ones.\\
\\
\section*{Discussions}
\hspace*{5mm}Fig.~\ref{fig:RER} shows $R_\mathrm{E}$(11) and $R_\mathrm{E}$(10), whose model sizes are shown in table.~\ref{table:table}, and $R$(11) and $R$(10) in 
fig.~\ref{fig:RT10RT11}. The frequency range of $R$(11) and $R$(10) is indicated within 0.338THz to 0.464 THz. These values deviate from BGX slightly owing to process and fabrication accuracy and so on.\\ 
\hspace*{5mm}The spot size's diameter of the incident wave in the measurement is about 20$a$ (6mm) \footnote{The sample size is 33$a$ $\times$ 33$a$ $\times$ 7$a$ in ref.~\cite{sakurai1}. }, which is much larger than the port size, 5$a$ $\times$ 5a in table.~\ref{table:table}. The scattering ratio to the measured reflected wave is able to be nearly neglected. Therefore, within BGX, the measured reflectivity is probably larger than the simulated one.\\
\hspace*{5mm}In fig.~\ref{fig:RER}(a), $R_\mathrm{E}$(11) and $R$(11) are approximately convex downward and similar spectra within BGX. Meanwhile, in fig.~\ref{fig:RER}(b), $R_\mathrm{E}$(10) and $R$(10) display concavity and convexity within PBG, which are characteristics in common. Within BGX, $R_\mathrm{E}$(11) and $R$(11) have smoother concavo-convex shape than $R_\mathrm{E}$(10) and $R$(10) do, relatively.
\begin{figure}[t]
\centering
\includegraphics[width=7.5cm,]{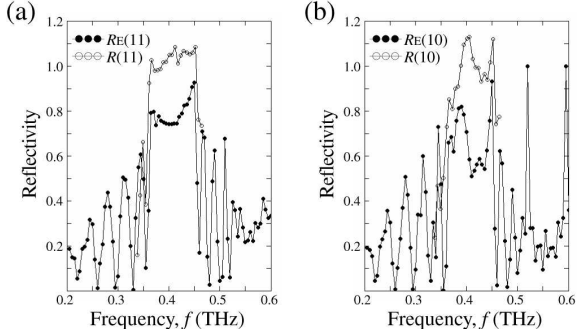}
\caption{\footnotesize Comparison of simulation reflected spectra (closed circle) with measured ones (open circle).}
\label{fig:RER}
\end{figure}
\section*{Conclusions}
\hspace*{5mm}The results of above discussions suggest that the polarization anisotropy of the measured reflected spectra does not apply to Maxwell's equations in appearance but apply to these equations in essential as expected. \\

\section*{Appendix}
\setcounter{figure}{0}
\renewcommand{\figurename}{Fig.~A.}
\hspace*{5mm}Some figures shown in ref.~\cite{sakurai1} and transmission measurement system in this work are collectively indicated in the appendix.\\ 
\begin{figure}[h]
\centering
\includegraphics[width=7cm, trim=0 0 0 -100]{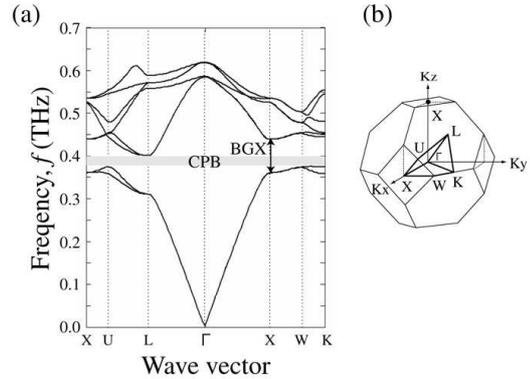}
\caption{\footnotesize (a) Calculated photonic band structure has complete photonic band gap, CPB (gray zone) at around 0.4 THz. BGX exists from 0.36 THz to 0.44 THz. (b) First Brillouin zone and reduced zone (heavy line) with high symmetry points.}
\label{fig:bandstructure}
\end{figure}
\begin{figure}[h]
\centering
\includegraphics[width=6.5cm, trim=0 0 0 150]{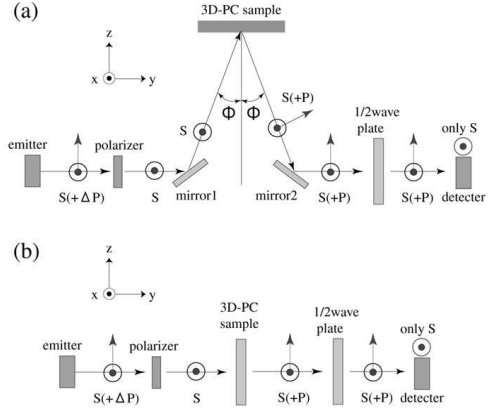}
\caption{\footnotesize Conceptual diagram of the measurement system with a THz-TDS. (a) Reflection measurement system. (b) Transmission measurement system. 
The polarization orientation of a THz wave from the emitter is S-p ∥ x-axis and the detector detects only S-p. Another polarization perpendicular to S-p is P-polarization (P-p) that is included in the incidence plane $\|$ z-axis. The designed 1/2 wave plate converts S-p into P-p and P-p into S-p at around 0.42 THz. This plate was used only in the case of P-p detection. S(+P) means S-p, P-p or the mixing of S-p and P-p. In fig.~A.\ref{fig:RTsystem}(a), the 3D-PC sample is so horizontally set that the layered direction is parallel to z-axis. It is parallel to Z-axis in fig.~\ref{fig:latticeA}. In this work, the 3D-PC sample was rotated in plane (001) instead of the S-p incident wave, relatively. The incident angle, $\phi$ is $7^{\ \circ}$, which is the normal incidence approximately.}
\label{fig:RTsystem}
\end{figure}
\begin{figure}[b]
\centering
\includegraphics[width=3cm, trim=0 0 0 0]{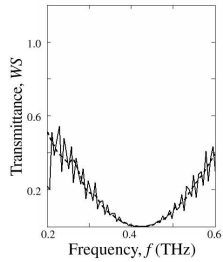}
\caption{\footnotesize Transmittance, $WS(f)$ of the designed 1/2 wave plate at around 0.42 THz.  For instance, S-p incident wave is nearly converted into P-p at 0.42 THz. The dash line includes the signal disposal from the back surface of the 1/2 wave plate. The solid line includes no disposal.}
\label{fig:WS}
\end{figure}
\begin{figure}[t]
\centering
\includegraphics[width=7cm, trim=0  0 0 1100]{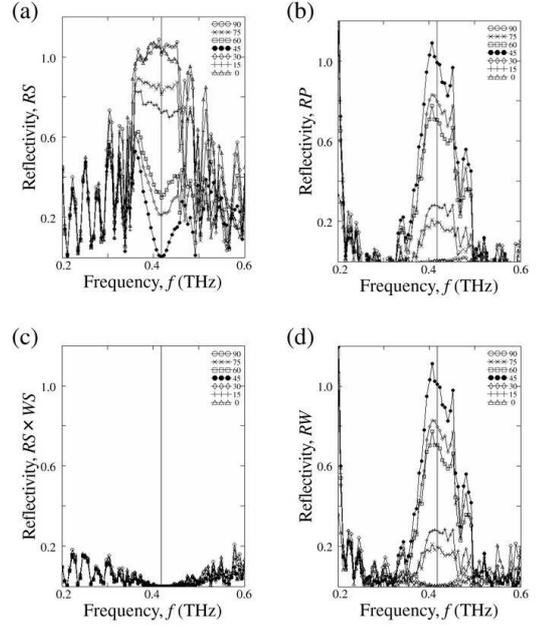}
\caption{\footnotesize Measured reflected spectra, $RS(f)$, $RP(f)$, $RS(f)$ $\times$ $WS(f)$ and $RW(f)$ from 0.2 THz to 0.6 THz for $\theta$ = 0$^\circ$ to 90$^\circ$ per 15$^\circ$.  $\theta$ is 3D-PC sample's rotation angle. The vertical solid line is 0.42 THz-line. $RS(f)$ is the S-p reflectivity with no 1/2 wave plate in fig.~A.\ref{fig:RSRPRW}(a), whose polarization orientation is parallel to that of the incident wave. $RS(f)$ is normalized by the Au refection spectra.  $RW(f)$ is the normalized reflectivity passing through the 1/2 wave plate. $RP(f)$ is the P-p reflectivity that is defined as $[RW(f)-RS(f) \times WS(f)]/[1-WS(f)]$.}
\label{fig:RSRPRW}
\end{figure}
\end{document}